# Self-Organizing Map and social networks: Unfolding online social popularity

A bottom-up sociological approach

Couronné Thomas, Beuscard Jean-Samuel, Chamayou Cédric
SENSE
Orange Labs
Paris, France
(Thomas.Couronne;  Jeansamuel.Beuscart; Cedric.Chamayou) at orange-ftgroup.com

*Abstract*— **The present study uses the Kohonen self organizing map (SOM) to represent the popularity patterns of Myspace music artists from their attributes on the platform and their position in the social network. The method is applied to cluster the profiles (the nodes of the social network) and the best friendship links (the edges). It shows that the SOM is an efficient tool to interpret the complex links between the audience and the influence of the musicians. It finally provides a robust *classifier* of the online social network behaviors.**

## I. INTRODUCTION

While the primary structure of the web was centered on data (web pages), the digitization of cultural goods, the emerging of new publishing technologies and the development of social networks have produced new forms of digital practices centered on user generated content (Herring [1], Huberman [2]). The success of social platforms of photo sharing (Flickr: Prieur *et al.* [3]), video (Youtube: Cha *et al.* [4]), music (Myspace: Caverlee [5]) raises new questions about the consumption of published contents. On these platforms, each user is able to manage its own visibility; he knows how many people viewed, commented, recommended, rated, forwarded his work, and his activity is strongly oriented by these ratings (Halavais [6]; Beuscart [7]). It is important to understand the patterns of these different ratings (does a high number of friends bring a high audience?) and what kind of content receives the most attention (are artists form major labels more listened?) (Anderson [8]).

Various indicators are available to estimate the popularity of a user depending on what we focus on. On one hand, we can assess popularity on the basis of the audience of the contents: number of visits of the producer's page, number of requests for the contents, number of comments. This is in fact the cumulative audience created by user's contents. On the other hand, when the popularity is focused on the user, we define it the influence or the authority, which reflects the number of people declaring the user as important for them, and measured by the social networks attributes like citation number, centrality, constrain level [9]. In this study, the word "influence" can be understood as network authority.

Audience indicators reflect the consumption of the digital goods, whereas influence describes the user's characteristics in the social network where content is shared. We demonstrate in [10] that theses two kinds of popularity are distinct and we try here to clarify the understanding of the relations between the two dimensions describing the online popularity. We can wonder if the users who have the highest audience have a specific position in the network and whether the audience elites (top charts) are the ones who have the highest influence in the network (best endogenous position). .

Social network analysis is a multi-disciplinary field including sociology, psychology and computer sciences, where tools and methods are quickly sophisticating. If many researchers are working on structural questions about network patterns ([11], [12]), few deal with the links between the online and offline attributes of a user and its network properties ([13], [14]). To better understand the mixed practices that the "user generated content" platforms engender, multidimensional analysis methods has to be used to process data about social network position and attributes. Because practices are not pre-defined and frequently mixed it is difficult to apply supervised classification tools or linear analysis. Therefore, our contribution to this complex data analysis is performed by the use of the Kohonen Self-Organizing maps (SOM) [15]. It provides an efficient way to distinguish various forms of popularity characteristics, producing a bi-dimensional visualization of multi-dimensional data which is strongly relevant for mixed social practices interpretations. It allows to unfold the hidden correlations between variables and is robust to extreme values. Because the SOM is a learning method, it builds a classification tool which finally allows performing successive temporal analysis and classifications.

Few works are dealing with social networks and Kohonen Self Organizing maps. Merelo-Guervos et al. [16] map blogs communities with SOM by a processing of a similarity matrix, which is defined by the network adjacency matrix (if A declares B as friend, then M(A,B)=1), where each feature vector is one column of the similarity matrix. They show that the SOM algorithm produces a relevant topological semantic map of the blogosphere.

Our work deals with the musical artist's popularity on MySpace social network aiming to compare the audience variables and the artist influence. MySpace is a social platform where musicians share musical content and information. For each artist, the platform displays his audience and rating scores: number of pages views –hits, number of comments, number of friends. The artist also declares some other users as "best friends"; these links are considered on the platform as a recommendation: choosing someone as "best friend" means, most of the time, strongly recommending his music [7].

In this paper we first present the Kohonen Self organized Maps principles. Then the data set elaboration is exposed. The first part of the analysis studies with the SOM algorithm the popularity characteristics of the profiles. The second part deals with the social network links clustering. Finally we observe how the links are distributed among the popularity clusters obtained and among the communities.

## II. KOHONEN SELF ORGANIZING MAP AND MULTIDIMENSIONAL CLUSTERING

A self organizing map is produced by a non supervised learning algorithm. It is composed by a set of n-dimension cells, organized in an m*p bi-dimensional surface. The basic version of this algorithm defines the cells number as $\sqrt{k}$ where k is the population to be clustered and m*p=sqrt(k). It differs from the "Growing self organizing map" [17] in which the number of cells varies depending on a overlapped structure.

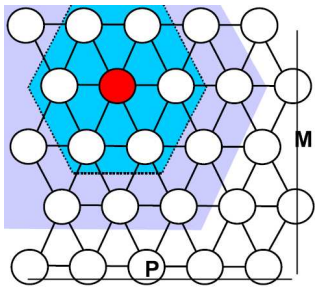

Figure 1. Schematic Self Organising Map

Each cell has a hexagonal shape (Fig. 1), therefore surrounded by six neighbors. Each individual of the population is characterized by a feature vector V of dimension n, where $V_t(i)$ is the value of the t-th variable among the n variables characterizing the individual i.

The first stage of the algorithm is the learning. The n dimensions of the m*p cells are randomly initialized. Then a subset of the population to model is randomly selected and individually processed by the SOM, moving from cell to cell until finding the "winner" cell : the one for which the feature vector is the closest. The feature vector of the winner cell is updated, taking into account the features' values of the individual. The new values of the cells' feature vector are smoothly broadcasted to neighbors' cells, to reduce the vectors gradient. The second stage of the algorithm is processing the global population to model. The last stage is to cluster the cells using similarity of their feature vector.

The Self organizing map thus provides a set of cells in a bi-dimensional plane, where each individual is associated with only one cell. Therefore two individuals with a similar feature vector will be associated with the same cell and the ones with opposed feature vector will have a topologically opposed position on the map. As said before, this method is applied to the MySpace music population.

## III. DATA CONSTRUCTION

We want to compare the audience and the influence characteristics in the Myspace social network. For this, we construct a crawler to collect a sample of this population.

Seven initial profiles are chosen among the French MySpace music top audience, and our crawler collects the profile data via the "best friendship" links. The number of best friends varies between 1 and 40. Our bread first search crawl consists in navigating through the best friend's links of the parent nodes during 3 iterations. This kind of crawling produces a sample with a relevant structure (good fitting of the clustering, density, and centrality values) but underestimates the in-degree and over-estimates the out-degree ([11],[18]). The information about audience is extracted for each profile and a sample of the influence network is built on the basis of the best friendship declaration. It provides an estimation of influence for each profile. Nevertheless, this network mixes the fans and the artists, but the best friendship link between fan and artist might not be semantically equivalent as between two artists. We chose, in this current study, to remove from the network all the non-artistic profiles but we record the in-degree value of each artist in the whole network.

In order to verify that the sample is not unusual, we have made several networks crawls from best friends to best friends, varying the entry numbers (from 3 to 10) and the parsing depth (from 2 to 4) (see table I).

If the number of nodes and the music profiles proportion (on MySpace the account's profile may be defined as "member" or "musician") depends on the crawling parameters, this ratio is around 50%. Next, for each crawl, a correlation test is applied to the four quantitative variables: comments, friends, in-degree and hits. Then a Mantel test is performed on the four correlation table, showing that the coefficients are significantly similar, i.e. the variables of each sample are correlated in the same proportions (Table II).

TABLE I. PROPERTIES OF FOUR NETWORK'S CRAWLS

| Network ID | depth | Entry |
|---|---|---|
| A | 3 | 4 |
| B | 3 | 7 |
| C | 2 | 10 |
| D | 4 | 3 |

The next table shows the Mantel Test between the crawl we study (B) and three other crawls (A,C, D).

TABLE II. MANTEL TEST: CORRELATION COEFFICIENTS

| MANTEL Test | R | p-value (bilateral) |
|---|---|---|
| B,A | 0,997 | 0,001 |
| B,C | 0,998 | 0,001 |
| B,D | 0,995 | 0,001 |

The data properties of the crawl 'B' are summarized in the table III:

TABLE III. DATASET PROPERTIES

| Total number of profiles | 21153 |
|---|---|
| Artists profiles | 13936 |
| Total Links number | 143831 |
| Links between artists number | 83201 |
| Reciprocal links rate (if A declares B as best friend, then B is highly valuable to declare A as best friend) | 40.1% |
| "Major" labeled artists | 3422 |
| "Indie" labeled artists | 7069 |
| "without" labeled artists | 3445 |

## IV. POPULARITY SPACE : AUDIENCE & INFLUENCE

### A. Data

Each musician is modeled as a node in the influence network graph, where the in-degree indexes the number of artists having declared the node as best friend (influencer) and the out-degree indexes the number of artists that a node has declared best-friends. The following variables are chosen to model the artist distribution in the audience-influence space and to construct the feature vector used for the SOM clustering:

- Page views (hits): indicator for the artist's audience
- Number of comments
- In-degree on the whole network, i.e. the cumul of fan and artists best friends links directed to the artist
- In-degree on the artist network, indicating the influence weight of the artist
- Page rank [19]: influence of the artist taking into account the influence of the neighbors.
- Reciprocity rate: fraction of A's best friends which have declared A as best friend: cooperative behavior
- Label: whether the artist record label is declared as "Major" (=3), "Indie" (=2), or "Other" (=1)

The distribution law which fits the best with the networks data is generally a power-law or zipf (scale-free) [20]. When methods require normality conditions, that kind of distribution doesn't allow immediate processing of the data.

For the page view, comments and in-degrees variables, a lillifors test [21] is applied to distribution normality hypothesis. This hypothesis is rejected (p<0.01). It appears that distribution follows a power-law. Therefore a log transformation is used instead of the value itself for theses non-normally distributed variables.

Thus, each node x of the graph G is characterized by a feature vector $V=(v_1(i), v_2(i), \ldots v_7(i))$ where $v_t(i)$ is the value of the t-th variable for the node i.

### B. Self Organizing map results

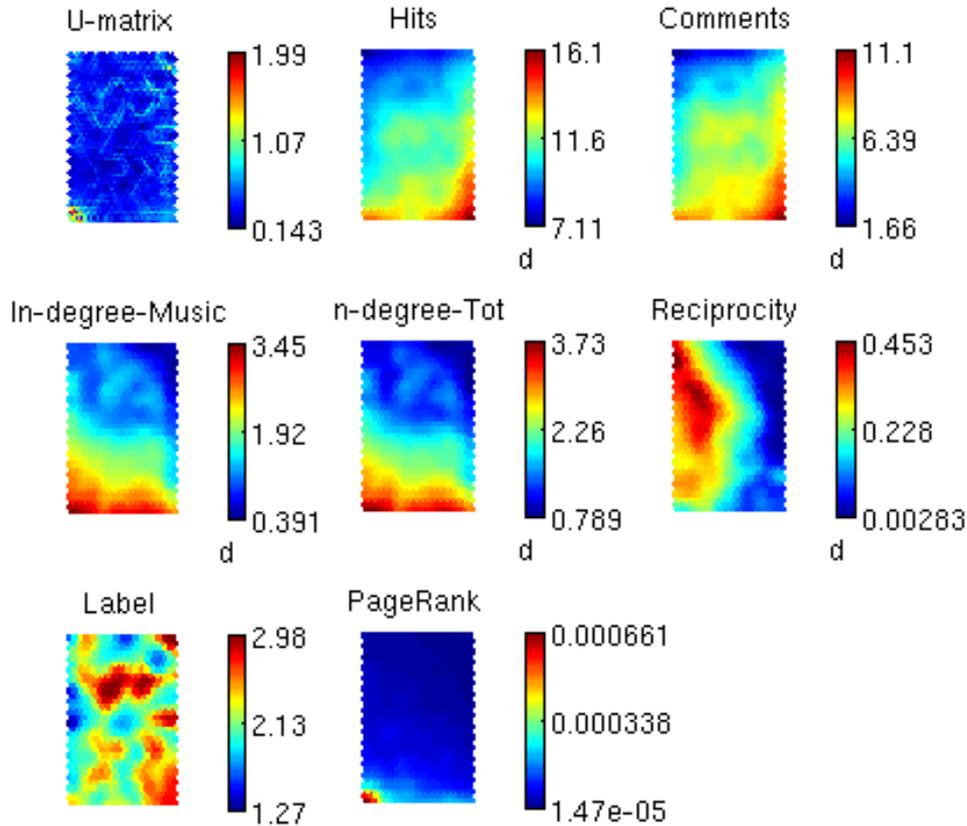

Figure 2. Self Organizing map of the nodes depending on their attributes and popularity properties

The Matlab version of the SOM algorithm used here is available on the Kohonen website.

The multi-dimensional processing of the nodes by the SOM provides the Fig. 2. Two complementary structures appear: north-south and east-west. The north-south axis describes the whole notoriety on both audience and influence: north we find the less famous artists, in the south the most famous artists. The east-west axis describes the reciprocity rate: the more the artists are to the west, the more they have reciprocal links. Nevertheless, if audience and influence are partly correlated and discriminate famous from anonymous artists, the trends are not exactly similar. Indeed, the south-west area is associated with the influential elites (highest in-degree and page-rank) and the south-east area is associated with the popular artists in terms of audience (highest page views and comments). If, logically, the audience elites are not without influence and influential elites are not without audience, the top artists of the audience and the influence do not overlap.

We can observe that the two kinds of in-degree distribution (whole network and artists sub-graph) are similar. It means that the distribution of best friendship between artists and between fan and artists is the same: the influential hierarchy appears to follow the same rules even if the links do not have the same semantic. The reciprocity distribution shows that the more the artists have a sociability dynamic (non null in-degree), the more the reciprocity is strong. But this reciprocity appears stronger for the middle and low level of influence than high influence ones: maybe because an influential artist can't have more than 40 best friends (and therefore can not cite everybody) or maybe because very influential artists are not linking back to people who link to them.

Finally we observe that the south-east area (audience elites) is associated with a strong presence of the « Major » labels. This label is also found in the one-third north part of the map, where these artists might have a social authority that the south east one majors' artists don't have.

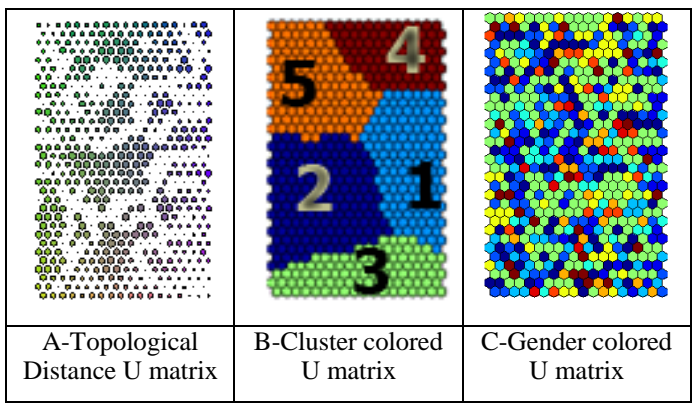

| A-Topological Distance U matrix | B-Cluster colored U matrix | C-Gender colored U matrix |

Figure 3. U matrix depending on distance, musical gender and clusters

A simple k-means clustering is performed on the cells distance matrix (Fig. 3-A). The expectation maximization algorithm is then employed to choose the best number of clusters. 5 clusters are defined here (Fig. 3-B):

- 1: Cyan: artists with very high audience, low influence and weak reciprocity links rate. They are Major labels' artists, with a strong offline marketing strategy that brings them a strong audience, but no intense MySpace strategy, therefore displaying relatively weak influence rates.

- 2: Dark blue: influential artists have a whole notoriety (audience + influence) lower than the superstars, but have an efficient innovative online marketing strategy that brings them an authoritative position on MySpace. They often correspond to trendy, avant-garde music.

- 3: Green are superstars, the most notorious artists, both influential and popular. They are the MySpace elites, having a high level online marketing strategy or a strong popularity in traditional Medias.

- 4: Brown are anonymous artist with a small audience and no active socializing practices.

- 5: Orange are socially dynamic artists with low audience. They are non professional artists swarms well inserted in the local music stages.

To know who belongs to the MySpace's elite, we focus on the artists belonging to the upper percentile of the page hits (audience) and the In-degree (influence).

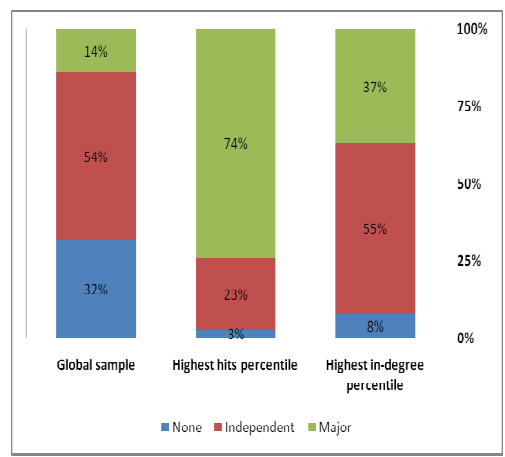

Figure 4. Label distribution for the global sample, the audience elits and the influence elits

Fig. 4 shows that while the 1% hit's elite is mainly associated with 'majors" artist, the most influential artists are a mix between independent and majors. If the artists labeled as "majors" are the undisputed stars of the audience on MySpace (they constitute 73% of the upper hit percentile, against 14% in the sample), the artists labeled as independent constitute an important part of the most influential artists (55% of the upper influential percentile, against 54% of the sample).

Fig. 3-C shows the SOM map where the cells color depends on the dominant music gender of the artists belonging to the cell. The map represents the proximity between the various music genres and the popularity, showing whether some musical genres are especially associated with some strong audience values or weak influence. The observation of the map does not confirm the hypothesis that some genres are associated with specific notoriety values. A complementary information which doesn't appear on the map and might help to read is that among the 10 dominant genders, 4 are a default musical declaration (alternative, null, other and indie) even though the artists can chose between 110 different genre.

## V. LINKS MODEL

In order to complete our understanding of the nature of the influence links, we study the characteristics of the links. Each link can be described by a feature vector, where each value is a gradient depending on the link emitter and receiver properties. The SOM processes the gradients, and shows if the anonymous artists declare as best friends elite' artists (strong or weak gradient). Let A be a variable describing the nodes, then gradient g is defined by:

g= (A(receiver)-A(emitter) )/(A(receiver)+A(emitter))

If the two nodes have the same value then g=0; if the emitter's value is highest, g tends to -1, otherwise 1.

The following variables are selected to build the feature vector for each link:

- Hits gradient
- Comments gradient
- Influence (In-degree) gradient
- Common predecessor in the graph (artists declaring commonly A and B as best friend)
- Common successors. (artists declared as best friend commonly to A and B)
- Whether the musical gender is similar (1) or differs (0)
- Whether the link is reciprocal or not (A->B and B<-A), (value=1 if reciprocal, 0 otherwise)

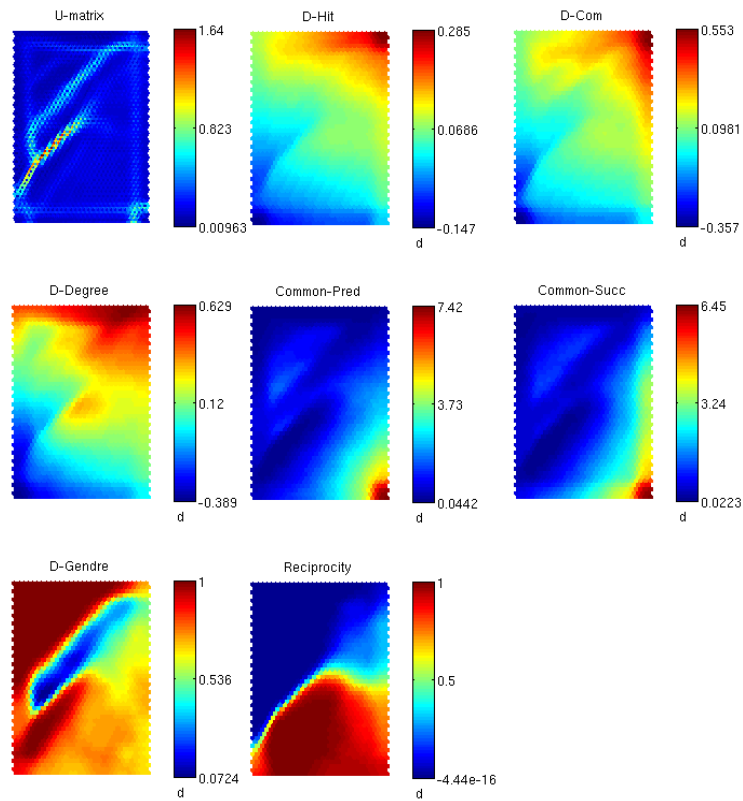

Figure 5. Self organizing map of the best friendship links

The processing of the influential links with a SOM shows (Fig. 5) that the main variable discriminating the links is reciprocity. Thus, the cells belonging to the north-west part of the map has non reciprocal links. This area is also associated with links between artists declaring different music genres: such kind of mixed genres links tend to be unidirectional. The north-east area is associated with links having a strong positive popularity gradient (audience and influence): the emitter of the influence link is less famous than the receiver; whereas the south-west area is associated with links having a strong negative gradient (the emitter is more famous than the receiver). This south-west area is also correlated with reciprocal links, while the north area has a weak reciprocity. These results show that the links with negative gradient are also reciprocal: the elite do not connect without reciprocity to a less-famous artist.

Finally, in the extreme south-west cells are the links with a strong similar neighborhood: they have a highest number of common predecessor and successor in the network. Thus, both variables are well correlated: when A and B are recommended by a common artist, then the probability that A and B recommend a common artist is strong. The second comment is that this co-influence area is associated with reciprocal links, medium level popularity and low to medium genre differences.

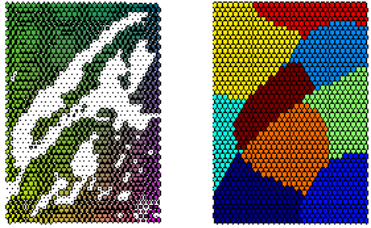

Figure 6.  A: Distance U matrix; B:clusters

The distance matrix (Fig. 6-A) shows the partition of the links. The links from the north-west part are non reciprocal; it links artists with a different musical genre and few common neighbors, and have a positive popularity gradient: they describe a cluster of relational links such as master - student or fan - star.

The links of the south-west part are reciprocal, with a common neighborhood; the popularity gradient is positive or negative, but with medium to low values: theses links can be understood as recommendation declarations.

After having observed independently the nodes and the edges properties, we compare the edges distribution among the nodes' clusters.

## VI. NODES' POPULARITY CLUSTERS AND INFLUENCE LINKS

The popularity clustering of artists raises questions about the influence distribution among theses clusters. More especially we can wonder in which proportion the influence links are distributed intra- and inter-clusters and, about the inter-clusters links, which clusters are inter-linked. A principal component analysis recalls the relative properties of each one of the five nodes clusters on the Fig. 7.

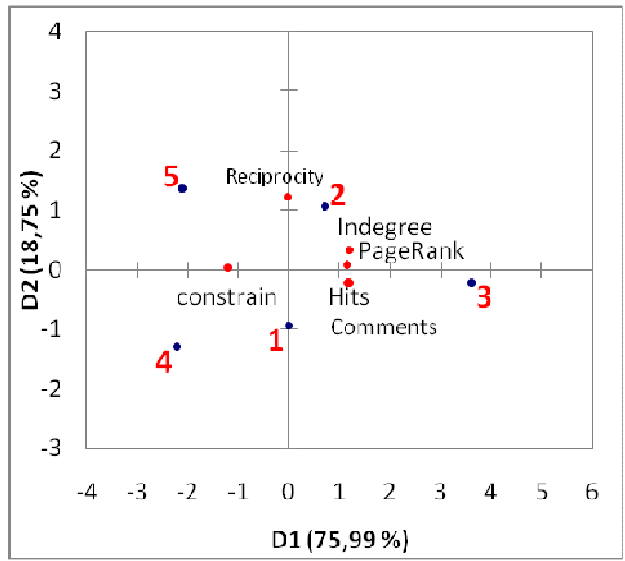

Figure 7.  Principal components analysis of nodes clusters (red: clusters IDs, nouns: quantitative variales

To clarify the interpretation we plot again the Self Organizing map with the clusters ids.

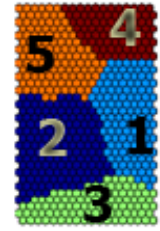

Figure 8.  Clusters labels

Let remember with help of the Fig. 7 and Fig. 8 that in the clusters 1 and 4 are artists having a part of audience and low influence. The 5th cluster gathers artists having a low influence and little audience. The cluster 2 is associated with influential artists, and the cluster 3 with globally popular artists.

We estimate the link frequency between clusters (best friendship declaration between artists who don't belong to the same cluster). Let M be the transition matrix: $M(i,j)$ is the number of links declared from artists of the i-th cluster to artists of the j-th cluster. The matrix M is normalized per line, i.e. defines the probability distribution $P(j/i)$ of the links emitted by the artist from the i-th cluster to the other j clusters where j=[1-5].

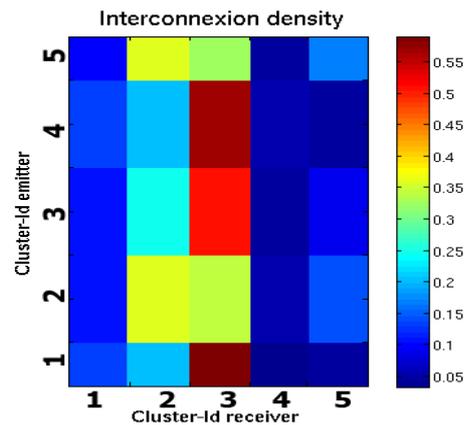

Figure 9.  Density matrix

The density matrix (Fig. 9) shows first of all that the most influential clusters are clusters 2 and 3 (highest linking probability). The cluster 3 possesses the most famous artists and the cluster 2 the artists having a strong social activity, with reciprocal links and community dynamics. The comparison of the emitter clusters shows that the clusters 2 and 5 have the same links destination trend, which differs to 4 and 1. The cluster 2 and 5 distribute their links to the clusters 2 and 3, with a small favor to the 2. The trend is inversed for the artists belonging to cluster 1 and 4, who declare as best friends mainly the most popular artists (cluster 3). Finally, the most popular artists (3) declare each others as best friend, but occasionally links to artists from cluster 2. These results confirm the following hypothesis:

- There is a social class of influential artists who, benefiting of the community dynamics, receives links from the whole population, elite included.
- The artists with links (small in degree reflecting a social activity) are at least equally likely to connect to influential artists as to notorious ones.
- The artists with a low influence and non reciprocal links tend to declare very famous artists as influence (and the elite non-reciprocal behavior explains the lowest level of their incoming links).
- The most popular artists tend to declare each other as influence.

## VII. SOCIAL NETWORK STRUCTURE ANALYSIS

Our data consists in a set of musician profiles (nodes) and best friendship links (edges). It is therefore relevant to draw the network and observe the macrostructure of the collected sample. The nodes' color is defined by the musical genre the artist belongs to; the nodes' size is defined by the artists' audience.

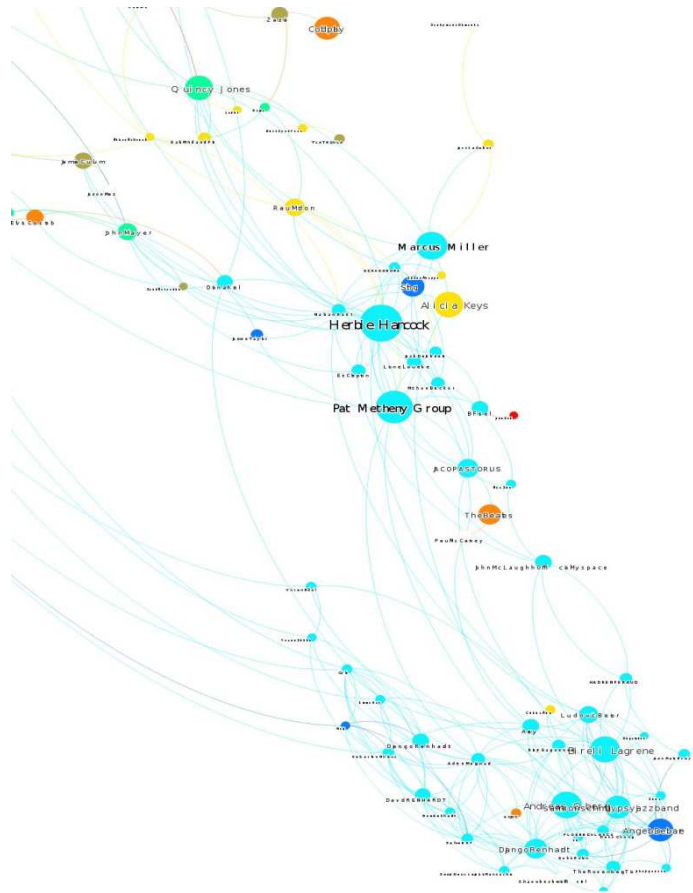
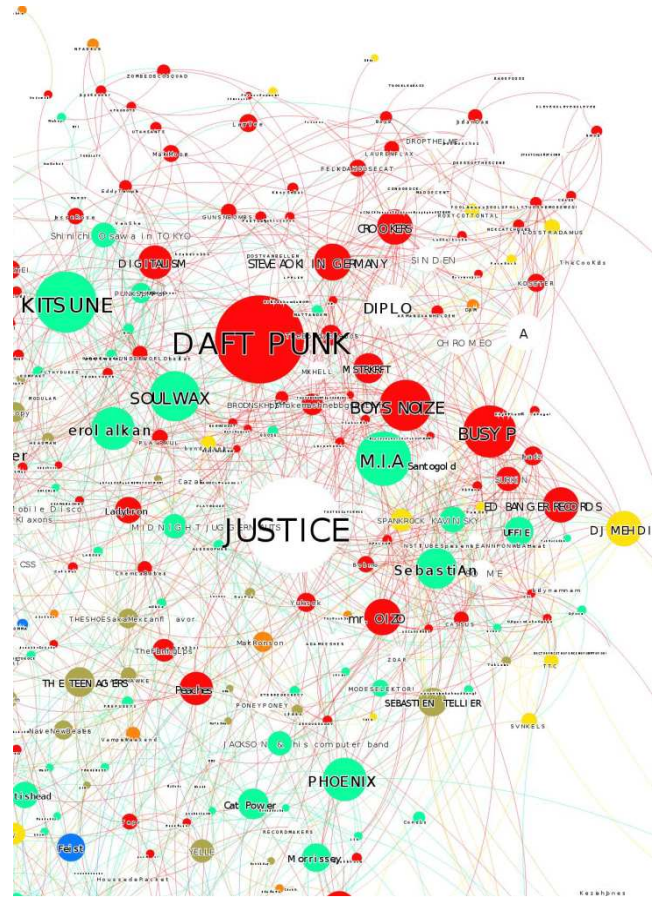

Figure 10. MySpace social network sample. The left one draws a clique of jazz musician; the right one shows the core of the influence network, composed by audience elites belonging to electro pop music, and surrounded by electro and indie artists

It appears (as shown on Fig.10) that some cliques (communities) are well identified and are composed by artists playing the same musical genre e.g. jazz or black music. These communities have a strong density and are linked to the main component by artists whose musical work is mixed or not well categorized like indie music. If the main component is composed by a mixture of musical genres, we observe that the audience' elite (biggest nodes) is well interlinked. Some commercial labels seem to have an efficient linking strategy (Ed Banger label): they are strongly central in the network, each artists citing each other. Moreover, they are well cited by artists having another musical genre. Because some artists of such community have a strong audience, people visiting these artists are highly likely to visit their friends: the audience is redirected to the other artists of the label.

In a second analysis, the network is processed with the "Linlog" algorithm [22]: after a spatial layout processing, each node obtains two dimensional spatial coordinates. Then a partitioning is applied on theses coordinates: artists having the same neighbors (predecessors or successors) are likely to belong in the same clique (same partition).

Each link is then either "intra clique" (both extremities in the same clique) or "inter clique".

We wonder if there is a relation between the kind of notoriety the artists have (as modeled by the clustering of the SOM) and the ability to link to artist who are not in the neighborhood.

Therefore, we compute for each SOM cluster the fraction of links emitted between artists who do not belong to the same spatial community.

TABLE IV. INTER CLIQUE LINKS FRACTION PER CLUSTER, DECREASING ORDER

| popularity cluster | 1 | 4 | 3 | 2 | 5 |
|---|---|---|---|---|---|
| link 'inter', fraction | 0,3276 | 0,2867 | 0,2291 | 0,2022 | 0,1869 |

The popularity clusters 1 and 4 are those who have the highest inter-community rate, as opposed to clusters 2 and 5.

It means that artists with audience and few or no influence are more able to recommend artists from various cliques and not exclusively their neighbors, than artists being in a reciprocal linking dynamic. It confirms the hypothesis that the artists of the second and fifth popularity cluster, who are defined as having a socialization activity, have a highest intra clique linkage.

VIII. CONCLUSION

The main idea of this work is to study the popularity of MySpace artists. More especially we want to detect whether or not the audience elites and the influential ones overlap. Therefore we introduce the Kohonen Self Organized Map to analyze the relations between audience (variables exogenous to the social network) reflecting the position in the music industry), and the influence (endogenous variable) revealing the macroscopic patterns of the online social structure. A SOM is established to cluster the nodes depending on their popularity variables. It confirms the hypothesis that audience and influence are two distinct dimensions. Then a SOM is performed on the links, which are characterized by the similarity of the nodes they link. It demonstrates that audience is not correlated to a reciprocal recommendation process. We then observe the inter-linkage between the popularity clusters, showing that elites tend to recommend each other. Finally, we process a network partitioning, to obtain best friendship cliques, and we show that the fraction of links between people of one clique is not the same depending on the popularity cluster of the artist who creates the recommendation link.

The SOM method appears robust to extreme values and stable: even if the initialization process is randomized, the obtained SOM is constant. One of the main interests of this method is its visualization mode, which seems strongly relevant to explore the non linear, hybrid, or mixed social practices. Moreover, the SOM learning properties provides an easy way to classify immediately new individuals, in the aim to observe the networks' dynamics among time.

If this study is only a contribution to the description of practices trends on MySpace, it contributes to a better understanding of the online popularity and suggests an efficient method to analyze the social network properties and the popularity characteristics.

The next step might be to establish topological maps of the social communities with the growing SOM version, to compare them to the diffusion of the social practices in the network.